# Koos Classification of Vestibular Schwannoma via Image Translation-Based Unsupervised Cross-Modality Domain Adaptation

Tao Yang[1][0000-0003-1440-0390] and Lisheng Wang[1][0000−0003−3234−7511]

[1] Department of Automation, Shanghai Jiao Tong University, Shanghai, China
{yangtao22,lswang}@sjtu.edu.cn

**Abstract.** The Koos grading scale is a classification system for vestibular schwannoma (VS) used to characterize the tumor and its effects on adjacent brain structures. The Koos classification captures many of the characteristics of treatment decisions and is often used to determine treatment plans. Although both contrast-enhanced T1 (ceT1) scanning and high-resolution T2 (hrT2) scanning can be used for Koos Classification, hrT2 scanning is gaining interest because of its higher safety and cost-effectiveness. However, in the absence of annotations for hrT2 scans, deep learning methods often inevitably suffer from performance degradation due to unsupervised learning. If ceT1 scans and their annotations can be used for unsupervised learning of hrT2 scans, the performance of Koos classification using unlabeled hrT2 scans will be greatly improved. In this regard, we propose an unsupervised cross-modality domain adaptation method based on image translation by transforming annotated ceT1 scans into hrT2 modality and using their annotations to achieve supervised learning of hrT2 modality. Then, the VS and 7 adjacent brain structures related to Koos classification in hrT2 scans were segmented. Finally, handcrafted features are extracted from the segmentation results, and Koos grade is classified using a random forest classifier. The proposed method received rank 1 on the Koos classification task of the Cross-Modality Domain Adaptation (cross-MoDA 2022) challenge, with Macro-Averaged Mean Absolute Error (MA-MAE) of 0.2148 for the validation set and 0.26 for the test set.

## 1 Introduction

Vestibular schwannoma (VS) is a benign, slow-growing tumor that occurs in the inner auditory canal from the inner ear to the brain [1]. The Koos grading scale is a classification system for vestibular schwannoma used to characterize the tumor and its effects on adjacent brain structures. Specifically, the Koos grading of VS is primarily determined by tumor size, location, and degree of compression of adjacent brain structures. The Koos classification captures many of the characteristics of treatment decisions and is often used to determine treatment plans. Specifically, the Koos grad-ing scale divides vestibular schwannoma into four grades according to criteria, as shown in **Fig. 1**. In recent years, with the advent of deep learning, it has become possible to improve patient outcomes and experience through standardization and personalization of VS treatment, while also significantly reducing physician workload [2]. So far, some deep learning-based VS automatic segmentation frameworks [3, 4] have been developed and demonstrated high segmentation accuracy on large real-world datasets. Meanwhile, a recent work [1] achieves automated classification of Koos rank through accurate segmentation of VS and adjacent brain structures.

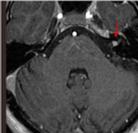

**Fig. 1.** The Koos grading scale with representative ceT1 and hrT2 images. Image courtesy to Kujawa [1].

However, Koos classification still faces challenges in practical applications, such as unsupervised, since medical data annotation is often time-consuming and expensive, and there is often the problem of domain shift between different imaging modalities. Unsupervised domain adaptation (UDA) has received much attention in the medical field because it does not require any additional annotation. However, the medical field lacks large benchmarks to evaluate the performance of UDA methods. crossMoDA is the first large multi-class benchmark for unsupervised cross-modality domain adaptation [5, 6]. The goal of the classification task of the crossMoDA 2022 challenge is to automate the Koos classification of VS from magnetic resonance imaging (MRI). The ceT1 scans are commonly used for Koos classification, but recent studies have shown the use of non-contrast imaging sequences, such as hrT2 imaging, can mitigate the risks associated with gadolinium-containing contrast agents. Furthermore, hrT2 imaging is more cost-efficient than ceT1 imaging. Therefore, the classification task of the crossMoDA 2022 challenge aims to automatically determine the Koos grade on unlabeled hrT2 scans using only labeled ceT1 scans based on the unsupervised domain adaptive approach. In this regard, we propose an unsupervised cross-modality domain adaptation method based on image translation by transforming annotated ceT1 scans into hrT2 modality and using their annotations to achieve supervised learning of hrT2 modality. Then, the VS and 7 adjacent brain structures related to Koos classification in hrT2 scans were segmented. Finally, handcrafted features are extracted from the segmentation results, and Koos grade is classified using a random forest classifier. The contributions of this paper are as follows:

- We propose an unsupervised cross-modality domain adaptation method for automatic Koos classification based on image translation.
- Image translation generates target modality images through image contrast transformation, thereby transferring the supervision information from the source domain into the target domain.
- The proposed method is validated on the challenging task of unsupervised cross-modality domain adaptation for Koos classification, outperforming other methods.

## 2 Related Work

In this section, we will present some work related to our proposed method, including Koos classification and unsupervised cross-modality domain adaptation.

### 2.1 Koos Classification

Koos classification system has been demonstrated to be a reliable method for vestibular schwannoma classification [7]. However, only recently have the first machine learning frameworks [1] for automatic Koos classification emerged. Before this, Koos grades of vestibular schwannoma could only be hand-labeled by neurosurgeons. Specifically, Kujawa [1] proposed a two-stage approach that implements classification after an initial segmentation stage. In the first stage, the VS and important adjacent brain structures are segmented. The segmentation annotations of these brain structures are obtained by the geodesic information flows (GIF) algorithm [8] rather than by hand. In the second stage, two complementary methods are used to perform Koos classification. The first method directly uses the dense convolutional network (DenseNet) [9] to classify the segmentation results of the first stage. The second method further extracts handcrafted features from the segmentation and then uses a random forest [10] for classification. Experimental results on a large dataset show that the performance of the method is comparable to that of neurosurgeons [1]. However, this method is based on supervised learning and lacks the ability of cross-modality domain adaptation for the unsupervised Koos classification task.

### 2.2 Unsupervised Cross-modality Domain Adaptation

Unlike natural images, medical images often have multiple complementary but heterogeneous modalities. Using the annotation information of one modality to help another modality build a task model can effectively reduce the annotation cost. Therefore, to bridge the differences between different imaging modalities, many cross-modality adaptation methods have emerged. Yang [11] proposed to find a shared content space through disentangled representations, enabling cross-modality domain adaptation between computed tomography (CT) and MRI images. The method embeds images from each domain into two spaces, a shared domain-invariant content space, and a domain-specific style space, and then performs tasks with representations in the content space. Chen [12] proposed synergistic image and feature alignment (SIFA), an unsupervised domain adaptation framework between MRI and CT images for cardiac substructure segmentation and abdominal multi-organ segmentation. However, these unsupervised cross-modality domain adaptation methods did not aim at our topic of the adaptation between ceT1 and hrT2 modalities.

For the adaptation between ceT1 and hrT2 modalities, Shin [13] proposed a self-training based unsupervised domain adaptation framework (COSMOS) for 3D medical image segmentation and validate it with automatic segmentation of VS and cochlea. The COSMOS realizes the modality transformation of the image through the target-aware contrast conversion network, while preserving the task features in the image during the transformation process. In addition, the method utilizes self-training [14] to iteratively improve segmentation performance. For the same task, Dong [15] proposed an unsupervised cross-modality domain adaptation approach based on pixel alignment and self-training (PAST). During training, pixel alignment is applied to transfer ceT1 scans to hrT2 modality to reduce the domain shift. Besides, Choi [16] proposed a domain adaptation method based on out-of-the-box deep learning frameworks for image translation and segmentation. In this method, an unpaired image-to-image translation model (CUT) based on patch-wise contrastive learning and adversarial learning is used for cross-modality domain adaptation. These unsupervised cross-modality adaptation methods achieve good results on VS and cochlea segmentation tasks. However, these adaptation methods that focus on segmentation tasks cannot be directly applied to the Koos classification task.

## 3 Methods

Our method consists of two parts: image translation and Koos classification, and the overall framework is shown in **Fig. 2**.

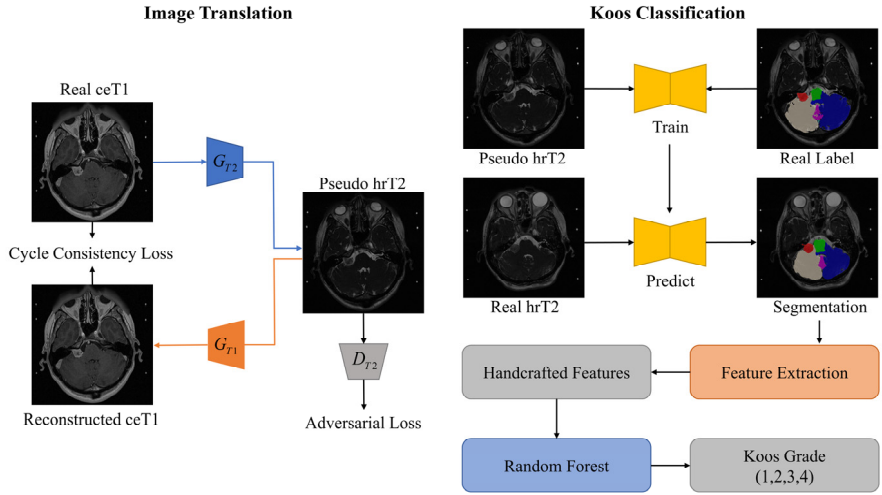

**Fig. 2.** The framework of Koos classification via image translation-based unsupervised cross-modality domain adaptation.

As shown in **Fig. 2**, image translation converts real ceT1 scans to the modality of hrT2 scans through an adversarial network, thus generating pseudo hrT2 scans. Koos classification consists of three steps. 1) Training a segmentation model in a supervised manner by using pseudo hrT2 scans and real labels. The trained segmentation model is used to predict the real hrT2 scans to generate brain structural segmentation. 2) Extracting handcrafted features from brain structure segmentation. 3) Training a random forest classifier for Koos grade prediction with handcrafted features extracted from the training set.

### 3.1 Image Translation

Since the source and target domain images in the training set provided by the cross-MoDA 2022 challenge are unpaired, we choose cycle-consistent adversarial networks (CycleGAN) [17] for image translation, i.e., we use CycleGAN to convert annotated ceT1 scans to hrT2 scans. CycleGAN achieved good results in the crossMoDA 2021 challenge [6, 13], which demonstrated the effectiveness of CycleGAN in bridging the gap between ceT1 scans and hrT2 scans. The typical CycleGAN is applied to 2D images, so all 3D images in the training set (including ceT1 scans and hrT2 scans) are sliced along the z-axis to obtain 2D images. Due to the different scanners used, the in-plane matrix sizes of the 3D images in the source and target domains do not match, which also leads to inconsistent sizes of the acquired 2D images. In this regard, we resize the image to a uniform size and do not crop the image so that CycleGAN has a global receptive field for the input image. In addition, considering the large computational effort of CycleGAN for image translation, we resize the original 2D image to the smallest size among all image sizes in the source and target domains. Specifically, the original 2D image is down-sampled to the smallest size among all image sizes in the source and target domains using the bicubic interpolation method. Then, the resized 2D images are fed directly into CycleGAN for training. According to previous work [15], the residual neural network (ResNet) was chosen as the generator of CycleGAN instead of U-net, while for the discriminator the default PatchGAN was chosen. After the training is completed, the annotated ceT1 scans can be translated into hrT2 scans using CycleGAN, thus training the segmentation model with the generated hrT2 scans in a supervised manner.

### 3.2 Koos Classification

We can train the segmentation model in a supervised manner by using the generated pseudo hrT2 scans and their corresponding annotations to be able to segment brain structures relevant to Koos classification in real hrT2 scans. According to previous work [1], there are 8 brain structures most relevant to Koos classification, including VS, pons, brainstem, cerebellar vermal lobules I-V, VI-VII, and VIII-X, left cerebellum (including the left cerebellum exterior and the left cerebellum white matter) and right cerebellum (including the right cerebellum exterior and the right cerebellum white matter). As in the previous work [16], we use the default 3D full resolution U-Net configuration of the nnU-Net [18] framework for training and inference for the brain structures segmentation task. The segmentation annotations of these brain structures are provided by the crossMoDA 2022 challenge and are initially obtained by the geodesic information flows (GIF) algorithm [8]. nnU-Net automates and condenses the critical decisions required to construct a successful segmentation pipeline for any given dataset [18]. Specifically, it is an out-of-the-box standardized segmentation framework that can self-configure the preprocessing, network architecture, and training pipeline for a given task without the need for manual intervention [13]. Therefore, we keep all automated configurations of nnU-Net without any modification to them. It is worth mentioning that, to reduce the large amount of training time caused by the nnU-Net's default 5-fold cross-validation, we did not use cross-validation, but used all the data to train a single segmentation model.

Eight brain structures and backgrounds are segmented on real hrT2 scans using the trained U-Net segmentation model. According to previous work [1], three handcrafted features are extracted for some structural segmentation masks and background masks, including volume, the shortest distance to the VS (DistVS), and contact surface with the VS (SurfVS). The left and right labels (left cerebellum and right cerebellum) are then transformed into ipsilateral and contralateral labels (ipsilateral cerebellum and contralateral cerebellum) related to the VS position to improve the classification performance. All the handcrafted features extracted are shown in **Table 1**. Based on these handcrafted features, a random forest classifier was trained on the training set and used to predict the Koos grade of patients.

**Table 1.** Handcrafted features extracted for Koos classification.

| Structures | Handcrafted features | | |
| --- | --- | --- | --- |
| | Volume | DistVS | SurfVS |
| Vestibular schwannoma (VS) | √ | | |
| Pons | | √ | |
| Brain stem | | √ | |
| Cerebellar vermal lobules I-V | | √ | |
| Cerebellar vermal lobules VI-VII | | √ | |
| Cerebellar vermal lobules VIII-X | | √ | |
| Ipsilateral cerebellum | | √ | |
| Contralateral cerebellum | | √ | |
| Background | | | √ |

## 4 Experiments

### 4.1 Dataset

All experimental data are provided by the crossMoDA 2022 challenge. The experimental data were obtained from two different hospitals, London hospital, and Tilburg hospital. For the London hospital, all images were obtained on a 32-channel Siemens Avanto 1.5T scanner using a Siemens single-channel head coil. For the Tilburg hospital, all images were obtained on a Philips Ingenia 1.5T scanner using a Philips quadrature head coil. A detailed description of the experimental data is shown in **Table 2**. All 3D images in the training set (including ceT1 scans and hrT2 scans) are sliced along the z-axis to obtain 2D images. Then, the 2D image is down-sampled to 256×256 using the bicubic interpolation method. Other than that, no other processing of the data is performed.

**Table 2.** The summary of data characteristics of the dataset.

| | Training set | | | | Validation set | | |
| --- | --- | --- | --- | --- | --- | --- | --- |
| | Source | | Target | | Target | | |
| Hospital | London | Tilburg | London | Tilburg | London | | Tilburg |
| Sequence | ceT1 | ceT1 | hrT2 | hrT2 | hrT2 | hrT2 | hrT2 | hrT2 |
| Number of scans | 105 | 105 | 83 | 22 | 105 | 28 | 4 | 32 |
| Annotation | √ | √ | × | × | × | × | × | × |
| In-plane matrix | 512×512 | 256×256 | 448×448 | 384×384 | 512×512 | 448×448 | 384×384 | 512×512 |

### 4.2 Evaluation Metrics

The macro-averaged mean absolute error (MA-MAE) [19] was used to evaluate the accuracy of the classification results. The MA-MAE is well-designed for ordinal and imbalanced classification problems and is defined as:

$$MA\text{-}MAE = \frac{1}{n}\sum_{j=1}^{m}\frac{1}{|T_j|}\sum_{x_i \in T_j}|P(x_i) - y_i|$$

where $n$ represents the number of all images, $T_j$ is the set of images with the true class label $y_j$, $P(x_i)$ and $y_i$ are the predicted class label and true class label of the image $x_i$, respectively.

### 4.3 Experimental Settings

**Image translation**: For CycleGAN, the weights of adversarial loss, cycle consistency loss, and identity loss are set to 1:10:5 by default. The training batch size is set to 10. The network was trained with Adam optimizer for 200 epochs, the first 100 epochs maintain an initial learning rate of 0.00015, and the latter 100 epochs decay linearly.

**Koos classification**: The settings for the random forest are default, but the number of trees is 100000, the maximum tree depth is 5, and the minimum number of samples per leaf is 2.

### 4.4 Experimental Results

All models were implemented in PyTorch 1.10 and trained and inference on an RTX 3090 GPU with 24GB of memory. The proposed method was trained on the training set (182 non-postoperative cases) and used to predict the Koos grade of patients in the validation and test sets. The proposed method received rank 1 on the Koos classification task of the crossMoDA 2022 challenge, with MA-MAE of 0.2148 for the validation set and 0.26 for the test set.

## 5 Discussion

Although our proposed method achieves the best results, this still falls short of the performance ($0.14 \pm 0.06$) [1] of fully-supervised random forest classification using hrT2 scans. Currently, the best Koos classification performance ($0.11 \pm 0.05$) in fully-supervised scenarios is already comparable to professional doctors ($0.11 \pm 0.08$) [1]. Therefore, the main bottleneck in unsupervised cross-modality Koos classification performance is the modality difference, the brain structures associated with Koos classification in ceT1 scans, including VS, are not fully converted to hrT2 modality. In addition, the degradation of segmentation performance caused by modality difference will also further increase the error of downstream Koos classification task. Therefore, it is more difficult for the unsupervised cross-modality Koos classification task to achieve fully-supervised accuracy than for the unsupervised cross-modality VS segmentation task. Fortunately, the unsupervised cross-modality segmentation performance of VS is already close to fully-supervised approaches [6, 13, 20]. Therefore, if the adaption performance of other brain structures related to Koos classification can be improved, the performance of unsupervised cross-modality Koos classification will likely be comparable to the fully-supervised approaches and professional doctors.

## 6 Conclusion

This paper proposed an image translation-based unsupervised cross-modality domain adaptation method for Koos classification of vestibular schwannoma. The classification performance on real datasets confirms that the proposed method has good cross-modality domain adaptability. In clinical practice, this domain adaptation method can effectively process unlabeled hrT2 modality images, thereby reducing the annotation cost and improving the diagnostic efficiency to a certain extent.

# References

1. A. Kujawa, R. Dorent, S. Connor, A. Oviedova, M. Okasha, D. Grishchuk, S. Ourselin, I. Paddick, N. Kitchen, T. Vercauteren, J. Shapey, Automated Koos Classification of Vestibular Schwannoma, Frontiers in Radiology 2 (2022).
2. J. Shapey, A. Kujawa, R. Dorent, S.R. Saeed, N. Kitchen, R. Obholzer, S. Ourselin, T. Vercauteren, N.W. Thomas, Artificial intelligence opportunities for vestibular schwannoma management using image segmentation and clinical decision tools, World Neurosurgery 149 (2021) 269-270.
3. J. Shapey, G. Wang, R. Dorent, A. Dimitriadis, W. Li, I. Paddick, N. Kitchen, S. Bisdas, S.R. Saeed, S. Ourselin, An artificial intelligence framework for automatic segmentation and volumetry of vestibular schwannomas from contrast-enhanced T1-weighted and high-resolution T2-weighted MRI, Journal Of Neurosurgery 134 (1) (2021) 171-179.
4. G. Wang, J. Shapey, W. Li, R. Dorent, A. Dimitriadis, S. Bisdas, I. Paddick, R. Bradford, S. Zhang, S. Ourselin, Automatic segmentation of vestibular schwannoma from T2-weighted MRI by deep spatial attention with hardness-weighted loss, in: International Conference on Medical Image Computing and Computer-Assisted Intervention, Springer, 2019, pp. 264-272.
5. J. Shapey, A. Kujawa, R. Dorent, G. Wang, A. Dimitriadis, D. Grishchuk, I. Paddick, N. Kitchen, R. Bradford, S.R. Saeed, S. Bisdas, S. Ourselin, T. Vercauteren, Segmentation of vestibular schwannoma from MRI, an open annotated dataset and baseline algorithm, Scientific Data 8 (1) (2021) 286.
6. R. Dorent, A. Kujawa, M. Ivory, S. Bakas, N. Rieke, S. Joutard, B. Glocker, J. Cardoso, M. Modat, K. Batmanghelich, A. Belkov, M.B. Calisto, J.W. Choi, B.M. Dawant, H. Dong, S. Escalera, Y. Fan, L. Hansen, M.P. Heinrich, S. Joshi, V. Kashtanova, H.G. Kim, S. Kondo, C.N. Kruse, S.K. Lai-Yuen, H. Li, H. Liu, B. Ly, I. Oguz, H. Shin, B. Shirokikh, Z. Su, G. Wang, J. Wu, Y. Xu, K. Yao, L. Zhang, S. Ourselin, J. Shapey, T. Vercauteren, CrossMoDA 2021 challenge: Benchmark of cross-modality domain adaptation techniques for vestibular schwannoma and cochlea segmentation, Medical Image Analysis (2022) 102628.
7. N.J. Erickson, P.G. Schmalz, B.S. Agee, M. Fort, B.C. Walters, B.M. McGrew, W.S. Fisher III, Koos classification of vestibular schwannomas: a reliability study, Neurosurgery 85 (3) (2019) 409-414.
8. M.J. Cardoso, M. Modat, R. Wolz, A. Melbourne, D. Cash, D. Rueckert, S. Ourselin, Geodesic information flows: spatially-variant graphs and their application to segmentation and fusion, IEEE transactions on medical imaging 34 (9) (2015) 1976-1988.
9. G. Huang, Z. Liu, L. Van Der Maaten, K.Q. Weinberger, Densely connected convolutional networks, in: Proceedings of the IEEE conference on computer vision and pattern recognition, 2017, pp. 4700-4708.
10. L. Breiman, Random forests, Machine learning 45 (1) (2001) 5-32.
11. J. Yang, N.C. Dvornek, F. Zhang, J. Chapiro, M. Lin, J.S. Duncan, Unsupervised domain adaptation via disentangled representations: Application to cross-modality liver segmentation, in: International Conference on Medical Image Computing and Computer-Assisted Intervention, Springer, 2019, pp. 255-263.
12. C. Chen, Q. Dou, H. Chen, J. Qin, P.A. Heng, Unsupervised bidirectional cross-modality adaptation via deeply synergistic image and feature alignment for medical image segmentation, IEEE transactions on medical imaging 39 (7) (2020) 2494-2505.

13. H. Shin, H. Kim, S. Kim, Y. Jun, T. Eo, D. Hwang, COSMOS: Cross-Modality Unsupervised Domain Adaptation for 3D Medical Image Segmentation based on Target-aware Domain Translation and Iterative Self-Training, arXiv preprint arXiv:2203.16557 (2022).

14. Q. Xie, M.-T. Luong, E. Hovy, Q.V. Le, Self-training with noisy student improves imagenet classification, in: Proceedings of the IEEE/CVF conference on computer vision and pattern recognition, 2020, pp. 10687-10698.

15. H. Dong, F. Yu, J. Zhao, B. Dong, L. Zhang, Unsupervised Domain Adaptation in Semantic Segmentation Based on Pixel Alignment and Self-Training, arXiv preprint arXiv:2109.14219 (2021).

16. J.W. Choi, Using Out-of-the-Box Frameworks for Unpaired Image Translation and Image Segmentation for the crossMoDA Challenge, arXiv preprint arXiv:2110.01607 (2021).

17. J.-Y. Zhu, T. Park, P. Isola, A.A. Efros, Unpaired image-to-image translation using cycle-consistent adversarial networks, in: Proceedings of the IEEE international conference on computer vision, 2017, pp. 2223-2232.

18. F. Isensee, P.F. Jaeger, S.A.A. Kohl, J. Petersen, K.H. Maier-Hein, nnU-Net: a self-configuring method for deep learning-based biomedical image segmentation, Nature Methods 18 (2) (2021) 203-211.

19. S. Baccianella, A. Esuli, F. Sebastiani, Evaluation measures for ordinal regression, in: 2009 Ninth international conference on intelligent systems design and applications, IEEE, 2009, pp. 283-287.

20. R. Dorent, S. Joutard, J. Shapey, S. Bisdas, N. Kitchen, R. Bradford, S. Saeed, M. Modat, S. Ourselin, T. Vercauteren, Scribble-based domain adaptation via co-segmentation, in: International Conference on Medical Image Computing and Computer-Assisted Intervention, Springer, 2020, pp. 479-489.